\documentclass[12pt]{book}
\usepackage[dvips]{graphicx,color}
\usepackage{makeidx,observe,cosmology}
\makeauthorindex
\makeindex

\BookTitle{New Trends in Theoretical and Observational Cosmology}
\CopyRight{\copyright 2001 by Universal Academy Press, Inc.}

\begin{document}

\BookTitle{\itshape New Trends in Theoretical and Observational
Cosmology} \CopyRight{\copyright 2001 by Universal Academy Press,
Inc.}

\pagenumbering{arabic}

\chapter{The Implications of $\bar{\nu}_e e^-\to W^-\gamma$
for the Detection of High-Energy $\bar{\nu}_e$}

\author{H. Athar$^{1,2}$ and
Guey-Lin Lin$^{2}$ \\
{$^{1}$\it Physics Division, National Center for Theoretical
Sciences, Hsinchu 300, Taiwan}\\
{$^{2}$\it Institute of Physics, National Chiao-Tung University,
Hsinchu 300, Taiwan}}

\AuthorContents{H. \ Athar and G.-L. \ Lin }
\AuthorIndex{Athar}{H.} \AuthorIndex{Lin}{G.-L.}

\section*{Abstract}
We discuss some motivations for detecting high-energy neutrinos
through the pure electroweak processes such as $\bar{\nu}_e e^-\to
W^- $ and $\bar{\nu}_e e^-\to W^-\gamma$. We argue that the latter
process can be viewed as an enhancement to the former one. The
event-rate enhancement is estimated.

\section{Introduction}
An observation of high-energy neutrinos beyond the atmospheric
background will mark the beginning of high-energy neutrino
astronomy. Several high-energy neutrino detectors, commonly known
as high-energy neutrino telescopes, are currently in an advanced
stage of their deployments. These detectors operate according to
the high-energy neutrino interactions, particularly those
occurring inside the high-energy neutrino telescopes. The studies
facilitated by the high-energy neutrino telescopes will complement
the high-energy gamma ray astronomy for understanding the origin
of cosmic high-energy radiation  \cite{Learned}.

In this talk, we shall discuss some motivations for detecting the
high-energy neutrinos through the pure electroweak interactions.
In particular, we focus on the resonant absorption process
$\bar{\nu}_{e}e^{-}\rightarrow W^{-} (\rightarrow
\bar{\nu}_{\mu}\mu^{-} $), now referred to as {\em Glashow
resonance} \cite{shelly,resonance} and the scattering process
$\bar{\nu}_{e}e^{-}\rightarrow W^{-}\gamma$ \cite{Brown}. The role
of the latter process will be discussed.

\section{The Detection Modes of High-Energy Neutrinos}
The typical detection of high-energy neutrinos is through the
charged-current neutrino-nucleon scattering
$\nu_l(\bar{\nu}_l)+N\to l^-(l^+)+X$. For $10^7 {\rm GeV} <
E_{\nu} < 10^{12} {\rm GeV}$, the cross section of this process
can be parameterized as $\sigma^{CC}(\nu N)= 5.5\cdot 10^{-36}{\rm
cm}^2(E_{\nu}/1 \ {\rm GeV})^{0.363}$ \cite{gqrs}. In general, the
cross section of neutrino-lepton scattering is suppressed except
for the $\bar{\nu}_e e^-$ scattering at the resonance energy
$E_{\bar{\nu}_{e}}=6.3\cdot 10^6 \ {\rm GeV}$. At this energy, the
resonant absorption $\bar{\nu}_e e^-\to W^-\to {\rm hadrons}$
occurs with a peak cross section $\sigma^{res}(s=m_W^2)=0.3 \
\mu$b. Although the resonant peak spreads only over the $W$-boson
width, the overall event rate in the high-energy neutrino
telescope is non-negligible after convoluting the absorption cross
section with the incoming  neutrino flux\cite{gqrs}.

It is interesting to note that the signature of $\bar{\nu}_e
e^-\to W^-\gamma$ is similar to that of the resonant absorption
provided the outgoing photon is soft. If the photon is hard, one
might be able to separate this process from the resonant
absorption by detecting the hard photon. There are a few reasons
for identifying the resonant absorption and $\bar{\nu}_e e^-\to
W^-\gamma$ processes in the high-energy neutrino telescopes. First
of all, unlike the neutrino-nucleon scattering, the resonant
absorption and $\bar{\nu}_e e^-\to W^-\gamma$ do not contain any
hadronic uncertainties. Therefore, they can be used to probe the
absolute flux of $\bar{\nu}_e$. Secondly, the $\bar{\nu}_e$ flux
measured in the high-energy neutrino telescope is in fact
identical to the intrinsic $\bar{\nu}_e$ flux from a
cosmologically  distant source, in spite of the flavor
oscillations taking place between the source and the high-energy
neutrino  telescope\cite{ajy}. Finally, we should point out that
only the $\bar{\nu}_e$ flux near the energy
$E_{\bar{\nu}_{e}}=6.3\cdot 10^6 \ {\rm GeV}$ can be probed by the
resonant absorption and $\bar{\nu}_e e^-\to W^-\gamma$. For other
neutrino energies, the cross section of neutrino-nucleon
scattering dominates.

\section{Results and Discussion}
The cross sections for resonant absorption and $\bar{\nu}_e e^-\to
W^-\gamma$ are both well known. We have further calculated the
photon spectrum of the latter process\cite{al}:
\begin{equation}\label{spectrum}
 \frac{\mbox{d}\sigma}{\mbox{d}y}=A(\lambda)
 \displaystyle{
 \left[(\lambda-1)(\lambda^2+1)y^{-1}+4\lambda^2(\lambda-1)y
 -2\lambda^3 y^2-
 \lambda(3\lambda^2-4\lambda+3)
 \right]},
\end{equation}
where $A(\lambda)=\sqrt{2}\alpha G_F/(\lambda^2(\lambda-1)^2)$,
with $\lambda=s/m_W^2$ and $y=E_{\gamma}/E_{\bar{\nu}_{e}}$. The
appearance of $1/y$ term in the spectrum
$\mbox{d}\sigma/\mbox{d}y$ indicates that the outgoing photon
tends to be soft. Indeed, we have computed $\langle y \rangle $,
the average value of $y$, for different incoming neutrino
energies. We have found $\langle y \rangle \simeq 1.3\cdot
10^{-3}$ for $E_{\bar{\nu}_{e}}=6.6 \cdot 10^{6}$ GeV, $\langle y
\rangle \simeq 6.0\cdot 10^{-3}$ for $E_{\bar{\nu}_{e}}=8.6 \cdot
10^{6}$ GeV, whereas $\langle y \rangle \simeq 9.7\cdot 10^{-2}$
for $E_{\bar{\nu}_{e}}=1.1 \cdot 10^{7}$ GeV. The corresponding
averaged photon energies in these cases are $8.6\cdot 10^3$ GeV,
$4.1\cdot 10^4$ GeV, and $10^6$ GeV respectively. With the above
energies, the photon radiation length is about $30$ cm in the
water or ice \cite{Stanev:1982au}. It turns out that the detection
of such photons is impractical given the large separation of
photomultiplier tubes in the current design of high-energy
neutrino telescopes.

Without detecting the outgoing photon in $\bar{\nu}_e e^-\to
W^-\gamma$, one should treat this process as an enhancement to the
resonant absorption. It is important to compare $\bar{\nu}_e
e^-\to W^-\gamma$ cross section with those of conventional
channels. The behaviors of these cross sections as functions of
$E_{\bar{\nu}_{e}}$ are shown in Fig. 1.
\begin{figure}[t]
  \begin{center}
    \includegraphics[height=17pc]{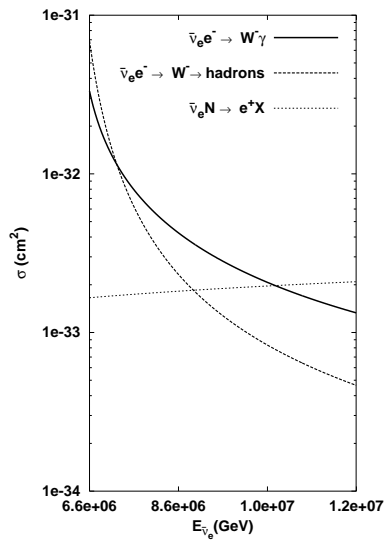}
  \end{center}
  \caption{High-energy $\bar{\nu}_{e}$
absorption cross section $\sigma$ (cm$^{2}$), over two different
target particles as a function of electron anti neutrino energy
$E_{\bar{\nu}_{e}}$ (GeV). The minimum value of
$E_{\bar{\nu}_{e}}$ corresponds to
$(M_{W}+\Gamma_{W})^{2}/2m_{e}$.}
\end{figure}
In this figure, we have included, besides the cross section of the
current process, the resonant cross section,
$\sigma_{\bar{\nu}_{e}e^{-}\rightarrow W^{-}\rightarrow
\mbox{hadrons}}$, taken from Ref. \cite{gaisser95}, as well as the
charged current deep-inelastic $\bar{\nu}_{e}$ scattering cross
section over nuclei, $\sigma^{CC}_{\bar{\nu}_{e}N\rightarrow
e^{+}X}$, taken from Ref. \cite{gqrs} with CTEQ4-DIS parton
distribution functions. From Fig. 1, we note that
$\sigma_{\bar{\nu}_{e}e^{-}\rightarrow W^{-}\gamma}$ dominates
over the other two for less than half an order of magnitude in
$E_{\bar{\nu}_{e}}$ ($7\cdot 10^{6}\leq
E_{\bar{\nu}_{e}}/\mbox{GeV}\leq 1\cdot 10^{7}$). The dominance is
however within a factor of 2.

In Ref. \cite{gqrs}, the hadron shower event rate due to the
Glashow resonance formation in the $\bar{\nu}_{e}e^-$ scattering
was estimated. We have computed an enhancement to this event rate
by incorporating the contribution of $\bar{\nu}_{e} e^-\to
W^-\gamma$ (see Fig. 1). The enhancement is typically about
$10\%$.  This enhancement, if become measurable, can be considered
as a signature of hard photon emission in the absorption process
$\bar{\nu}_{e}e^{-}\rightarrow W^{-}\gamma$ in a high-energy
neutrino telescope. In this estimate, we have used the
$\mbox{d}\displaystyle{N}/
\mbox{d}\displaystyle{E_{\bar{\nu}_{e}}}$ given by
\begin{equation}\label{flux}
 \frac{\mbox{d}N}{\mbox{d}E_{\bar{\nu}_{e}}}\simeq 10^{-8}
 \left(\frac{E_{\bar{\nu}_{e}}}{1 \mbox{GeV}}\right)^{-2}
 \mbox{cm}^{-2}\mbox{s}^{-1}\mbox{sr}^{-1}\mbox{GeV}^{-1},
\end{equation}
for the relevant $E_{\bar{\nu}_{e}}$ range. In Eq.~(\ref{flux}),
the $\mbox{d}\displaystyle{N}/
\mbox{d}\displaystyle{E_{\bar{\nu}_{e}}}$ is the downward going
differential $\bar{\nu}_{e}$ flux arriving at the high-energy
neutrino telescope. Briefly, it is for the gamma ray burst
fireball model proposed in Ref. \cite{waxman}, where $p\gamma $
interactions are suggested to produce the high-energy
$\bar{\nu}_{e}$ flux at the gamma ray burst fireball site.

In conclusion, we have provided some motivations for detecting the
high-energy $\bar{\nu}_e$ through the resonant absorption process
$\bar{\nu}_e e^-\to W^-$ and the scattering $\bar{\nu}_e e^-\to
W^-\gamma$, which are purely electroweak in nature. The latter
process is shown to enhance the event rate of the former process
by about $10\%$.

H.A. is supported by Physics Division of National Center for
Theoretical Sciences. G.L.L. is supported by the National Science
Council under the grant number NSC90-2112-M009-023 and Research
Excellence Project on Cosmology and Particle Astrophysics (CosPA)
funded by Taiwan's Ministry of Education.



\begin{thebibliography}{99}
\bibitem{Learned}
For recent review articles, see, for instance, J.~G.~Learned and
K.~Mannheim, Ann. Rev. Nucl. Part. Sci. 50 (2000) 679;
 F.~Halzen,
Phys. Rept. 333 (2000) 349
and references  therein.
%
\bibitem{shelly}
S. L. Glashow, Phys. Rev. 118 (1960) 316.
%
\bibitem{resonance}
J. N. Bahcall and S. C. Frautschi, Phys. Rev. 135B (1960) 788;
V.~S.~Berezinski\u{i} and A.~Z.~Gazizov, JETP Lett. 25 (1977) 254
[Pisma Zh. Eksp. Teor. Fiz. 25 (1977) 276];
 V.~S.~Berezinsky,
D.~Cline and D.~N.~Schramm, Phys. Lett. B78 (1978) 635;
I.~M.~Zheleznykh and \'{E}.~A.~Ta\u{i}nov, Sov. J. Nucl. Phys. 32
(1980) 242
 [Yad. Fiz. 32 (1980) 468];
V.~S.~Berezinsky and V. L. Ginzburg, Mon. Not. R. Astr. Soc. 194
(1981) 3;
 R.~W.~Brown and F.~W.~Stecker, Phys. Rev. D26 (1982)
373;
G.~Domokos and S.~Kovesi-Domokos, Phys. Lett. B346 (1995) 317.
%
\bibitem{Brown}
R.~W.~Brown, D.~Sahdev and K.~O.~Mikaelian, Phys. Rev. D20 (1979)
1164;
 K.~O.~Mikaelian and I.~M.~Zheleznykh, Phys. Rev. D22 (1980)
2122;
 V.~S.~Berezinski\u{i} and A.~Z.~Gazizov, Sov. J. Nucl. Phys.
33 (1981) 120  [Yad. Fiz. 33 (1981) 230];
 F.~Wilczek, Phys. Rev.
Lett. 55 (1985) 1252;
D.~Seckel, Phys. Rev. Lett. 80 (1998) 900.
%
\bibitem{gqrs}
R.~Gandhi, C.~Quigg, M.~H.~Reno and I.~Sarcevic, Phys. Rev. D58
(1998) 093009.
%
\bibitem{ajy}
H.~Athar, M.~Je\.{z}abek and O.~Yasuda, Phys. Rev. D62 (2000)
103007 and references  therein.
%
\bibitem{al}
H.~Athar and G.-L.~Lin, hep-ph/0108204.
%
\bibitem{Stanev:1982au}
 T.~Stanev, C.~Vankov, R.~E.~Streitmatter, R.~W.~Ellsworth and T.~Bowen,
Phys.\ Rev.\ D25 (1982) 1291.
%
\bibitem{gaisser95}
T.~K.~Gaisser, F.~Halzen and T.~Stanev, Phys. Rept. 258 (1995) 173
[Erratum-ibid. 271 (1996) 355].
%
\bibitem{waxman}
E.~Waxman and J.~Bahcall, Phys. Rev. Lett. 78 (1997) 2292.
%

\end{thebibliography}
\end{document}